\documentclass[a4,11pt]{cipro}
\usepackage{graphics}
\usepackage{graphicx}
\usepackage{amsmath,amsxtra,amssymb,latexsym, amscd,color}
\usepackage[mathscr]{eucal}
\usepackage[symbol]{footmisc}
\begin{document}­
\def\figup{\pagebreak$\quad$\vskip-0.7cm}
\def\Mr{\MakeUppercase}
\def\vsa{\vskip-0.2cm}
\def\vssa{\vskip-0.4cm}
\def\vs{\vskip0.2cm}
\def\vsm{\vskip0.1cm}
\def\vss{\vskip0.4cm}
\def\bce{\begin{center}}
\def\ece{\end{center}}
\def\n{\noindent}
\def\nn{\nonumber}
\def\ce{\centerline}
\def\disp{\displaystyle}
\def\abs{\begin{abstract}\rightskip-1.2cm\it}
\def\eabs{\end{abstract}}
\parindent=1.05cm
%TITLE OF THE ARTICLE%%%%%%%%%%%%%%%%%%%%%%%%%%%%%%%%%%%%%%%%%%%%%%%%%%%%
\title
{\large The principle of least action for fields containing higher
order derivatives$^+$ } \thanks {$^+$Submitted to the $33^{th}$
Vietnam National Conference on Theoretical Physics, from $4^{th}$-
$7^{th}$ August 2008, Danang, Vietnam. }
  %Ngay nhan bai
\maketitle
%SHORTEN NAME OF AUTHORS AND TITLE OF THE ARTICLE%%%%%%%%%%%%%%%%%%%%%%%%%%%%%%%%
\markboth{NGUYEN DUC MINH}{THE PRINCIPLE OF LEAST ACTION}
%AUTHORS AND ADDRESSES %%%%%%%%%%%%%%%%%%%%%%%%%%%%%%%%%%%%%%%%%%%%%%%%%
\begin{center}
NGUYEN DUC MINH, \footnote{E-mail: nguyenducminh$3\_1$@yahoo.com} \\
{\it
Department of Physics, College of Science,VNU\\
}
\end{center}
%ABSTRACT %%%%%%%%%%%%%%%%%%%%%%%%%%%%%%%%%%%%%%%%%%%%%%%%%%%%%%%%%%
\abs   %Phan Abstract
When generalizing the principle of least action for fields
containing higher order derivatives, in general, it is not possible
not to take into account the surface integrated term since it gives
direct contribution to the forms of the equations of motion, of the
energy-momentum tensor and of the angular-momentum tensor. This
result is applied to two examples. In two dimensional gravity, it is
essential to supplement Dirichclet condition for the surface
integrated terms to vanish. On the contrary, in Hamiltonian
description of the open rigid string, the equation of motion is
modified by surface integrated terms. Boundary conditions are
classified according to forms of certain equations of motion  \eabs

\section{INTRODUCTION}

Dynamical systems described in terms of higher-order Lagrangians
have exhibited a lot of interesting aspects in connection with the
gauge theories \cite{ref1}, gravity \cite{ref2,ref3}, supersymmetry
\cite{ref4,ref5}, string models \cite{ref6,ref7}, and other problems
\cite{ref8,ref9,ref10}. However, up to now it appears difficult to
generalize the principle of least action for fields containing
higher order derivatives due to the existence of surface integrated
components \cite{ref11,ref12,ref13}.

The purpose of this paper is to apply different boundary conditions
to consider the role of surface integrated terms, find the correct
forms of the equations of motion, the energy-momentum tensor and the
angular-momentum tensor in the appearance of boundary conditions.

The paper is organized as follows. Section II presents the
derivation of energy-momentum tensor, angular-momentum tensor and
study the influence of boundary conditions on Noether theorem.
Section III is devoted to $2$D gravity. In section IV, the
corresponding formulae in the case of string theory are given.
Section V is for the drawn conclusion.

\section{FIELD SYSTEM AND BOUNDARY CONDITIONS}

\subsection{The principle of least action}

Let us consider Lagrangian density for a free scalar field, which
contains second order dervivatives
\begin{equation} \label{1}
L\left( {\varphi ,\partial _\mu  \varphi ,\partial _\mu  \partial
_\nu  \varphi } \right).
\end{equation}

The equation of motion is derived by making an infinitesimal
variation $\delta \varphi(x)$, and requiring the corresponding
variation of the action to vanish
\begin{equation} \label{2}
0 = \delta \int\limits_\Omega  {L\,d^4 x}  \hfill.
\end{equation}
By denoting
\[
L_\varphi   = \frac{{\partial L}} {{\partial \varphi }} - \partial
_\mu  \frac{{\partial L}} {{\partial \varphi _{,\mu } }} + \partial
_\nu  \partial _\mu  \frac{{\partial L}} {{\partial \varphi _{,\mu
\nu } }},\,\,\,A_\mu   = \left( {\partial _\nu  \frac{{\partial L}}
{{\partial \varphi _{,\mu \nu } }} - \frac{{\partial L}} {{\partial
\varphi _{,\mu } }}} \right)\bar \delta \varphi  - \frac{{\partial
L}} {{\partial \varphi _{,\mu \nu } }}\bar \delta \varphi _{,\nu },
\]
and integrating by parts (\ref{2}), we can rewrite the principle of
least action in the form
\begin{equation} \label{3}
0 = \int\limits_\Omega  {d^4 xL_\varphi  }  + \int\limits_{\partial
\Omega } {\left( { - A_\mu   + L\delta x_\mu  } \right)} d\sigma
_\mu.
\end{equation}

Ignoring surface integrated term in (\ref{3}), the corresponding
Euler-Lagrange equation will be
\begin{equation} \label{4}
L_\varphi   = 0\,\,\,\,or\,\,\,\,\,\frac{{\partial L}} {{\partial
\varphi }} - \partial _\mu  \frac{{\partial L}} {{\partial \varphi
_{,\mu } }} + \partial _\nu  \partial _\mu  \frac{{\partial L}}
{{\partial \varphi _{,\mu \nu } }} = 0.
\end{equation}

Let consider a space and time translation in which the variation of
coordinate is ${\delta x_\mu   = \varepsilon _\mu  }$ and the field
function transformation is
\begin{equation} \label{5}
\varphi \left( {x_\mu  } \right)\,\, \to \,\,\varphi '\left( {x_\mu
} \right) = \varphi \left( {x_\mu  } \right) - \varepsilon _\mu
\partial _\mu  \varphi _\mu.
\end{equation}

The substitution of relation (\ref{5}), (\ref{4}) into expression
(\ref{3}) yields to the following local conservation law
\begin{equation} \label{6}
T_{\rho \mu }  =  - \frac{{\partial L}} {{\partial \varphi _{,\mu }
}}\varphi _{,\rho }  - \frac{{\partial L}} {{\partial \varphi _{,\mu
\nu } }}\varphi _{,\rho \nu }  + \partial _\nu  \frac{{\partial L}}
{{\partial \varphi _{,\mu \nu } }}\varphi _{,\rho }  + L\delta _{\mu
\rho } ,\,\,\,\partial _\mu  T_{\rho \mu }  = 0.
\end{equation}

$T_{\rho \mu } $ is the energy-momentum tensor of the system whose
Lagrangian contains second order derivatives. From its local
conservation we get four constants of motion
\begin{equation}  \label{7}
P_\nu   = -\int {d^3x } T_{\nu 0}.
\end{equation}

The last case we will consider is the invariance with respect to
Lorentz transformations
\begin{equation}  \label{8}
x_\mu   \to x'_\mu   = x_\mu   + \varepsilon _{\mu \nu } x_\nu,
\end{equation}
\begin{equation}  \label{9}
\psi _\alpha  (x) \to \psi '_\alpha  (x') = \psi _\alpha  (x) +
\frac{1} {2}\varepsilon _{\mu \nu } (\Sigma _{\mu \nu } )_{\alpha
\beta } \psi _\beta  (x).
\end{equation}

Using (\ref{8}), (\ref{9}), (\ref{4}) we find that the expression
(\ref{3}) gives
\begin{equation}  \label{10}
\frac{1} {2}\varepsilon _{\mu \rho } \int\limits_{\partial \Omega }
{M_{\mu \nu \rho } d\sigma _\mu  }  = 0,
\end{equation}
where
\begin{equation}  \label{11}
\begin{array}{l}
 M_{\mu \rho \sigma }  = x_\mu  \left( {T_{\rho \sigma }  - \psi
_{\alpha ,\rho } \delta _{\rho \nu }\frac{{\partial L}}{\partial
\psi_{\alpha ,\sigma \nu}}}\right)- x_\rho  \left( T_{\mu \sigma } -
\psi _{\alpha ,\mu } \delta _{\rho \nu }\frac{{\partial L}}{\partial
\psi_{\alpha ,\sigma \nu}}\right)+\\
  \,\,\,\,\,\,\,\,\,\,\,\,\,\,\,\,\,\,\,\,\, + \left(
{\frac{{\partial L}}{{\partial \psi _{\alpha ,\sigma } }} -
\partial _\nu  \frac{{\partial L}}{{\partial \psi _{\alpha ,\sigma
\nu } }}} \right)(\Sigma _{\mu \nu } )_{\alpha \beta } \psi _\beta -
\frac{{\partial L}}{{\partial \psi _{\alpha  ,\sigma \nu } }}(\Sigma
_{\mu \rho } )_{\alpha \beta } \psi _{\beta ,\nu }.\\
 \end{array}
\end{equation}

It is clear that $ M_{\mu \rho }  =  - \int\limits_{\partial \Omega
} {M_{\mu \nu 0} {\mathop{\rm d}\nolimits} ^3 x} $ is a 2nd rank
tensor. This quantity is the angular-momentum tensor of the field.
These results (\ref{6}), (\ref{11}) are agreed with the formulae in
\cite{ref14} if we put $n=2$.

\subsection{Boundary conditions}

In the above subsection, we see that conservation laws associated
with energy-momentum tensor and angular-momentum tensor (Noether
theorem) is only correct if field function $\varphi$ satisfies the
equations of motion. This happens when the surface integrated term
in (\ref{3}) is vanished. Nevertheless, in the case of Lagrangian
with second order derivatives,  surface integrated components do not
vanish naturally. Thus, to Noether theorem is satisfied, certain
boundary conditions are necessary.

Now, we shall discuss more carefully about one of the conservation
laws of Noether theorem - the  energy conservation. When Lagrangian
L contains second order derivatives, there are two independent
configuration-space-type variables $\varphi $, ${\dot \varphi }$ and
the corresponding canonical momenta
\begin{equation} \label{12}
\pi  = \frac{{\partial L}} {{\partial \dot \varphi }} -
\frac{\partial } {{\partial t}}\frac{{\partial L}} {{\partial \left(
{\ddot \varphi } \right)}} - \partial _i \frac{{\partial L}}
{{\partial \left( {\partial _i \dot \varphi } \right)}},\,\,\,\,\,s
= \frac{{\partial L}} {{\partial \left( {\ddot \varphi }
\right)}}\,\,(i = 1,2,3).
\end{equation}

Substituting (\ref{12}) into (\ref{7}), the conserved energy $P_0$
will be
\begin{equation} \label{13}
\,P_0 \,\,\,\, = \int {d_{}^3 } x\left( {\pi \dot \varphi  +
s\,\ddot \varphi  - L} \right) + \frac{1} {2}\int {d_{}^3 }
x\partial _i \left( {\dot \varphi \frac{{\partial L}} {{\partial
\left( {\partial _i \dot \varphi } \right)}}} \right).
\end{equation}

In other words, Hamilton function of the system is
\begin{equation} \label{14}
H\, = \int {d_{}^3 } x\left( {\pi \dot \varphi  + s\,\ddot \varphi -
L} \right).
\end{equation}

Therefore, from (\ref{13}), (\ref{14}) we see that, unlike in  the
case of field containing first order derivatives, when the second
order appears, Hamilton function $H$ differs from conserved energy
$P_0$ a surface integrated term
\begin{equation} \label{15}
M = \int\limits_R {Kd^3 x\, = } \frac{1} {2}\int\limits_R {\partial
_i \left( {\dot \varphi \frac{{\partial L}} {{\partial \left(
{\partial _i \dot \varphi } \right)}}} \right)d^3 x\,}.
\end{equation}

This quantity does not vanish in general, but depends on the
boundary conditions. Its contribution to the energy formula will be
considered more below.

In conclusion, since the surface integrated terms must be taken
account, we must supplement boundary conditions for Noether theorem
to be correct with higher order field. Upon the choice of boundary
conditions, some changes will appear in the theorem. This will be
discussed in the next two sections.

\section{TWO DIMENSIONAL GRAVITY}

We have shown that with high order Lagrangian, the contribution of
surface integrated terms depends on boundary conditions. In this
section, an appropriate boundary condition for these terms in $2$D
gravity to vanish will be derived.

In Teitlboim model \cite{ref15}, $2$D  gravity has Lagrangian
\begin{equation} \label{16}
L = \sqrt { - g} \,k^{ - 1} \left( {R - \Lambda } \right).
\end{equation}

By using $ \left\| {g_{\mu \nu } } \right\| = e^\varphi  \left(
{\begin{array}{*{20}c}
   {(\eta ^1 )^2  - (\eta ^ \bot  )^2 } & {\eta ^1 }  \\
   {\eta ^1 } & 1  \\
\end{array}} \right)$, the above Lagrangian can be rewritten in the
form
\begin{equation} \label{17}
L = \frac{1}{{\eta ^ \bot  }}\left( {\dot \varphi  - \varphi '\eta -
2\eta ^{1'} } \right)^2  - \eta ^ \bot  \varphi '^2  + 4\eta ^ \bot
\varphi '' + 2\eta ^ \bot  \Lambda \,e^\varphi,
\end{equation}
where only scale conformal factor $\varphi$ is dynamical variable ($
\varphi (t,x),\,\,x_\mu   = (x_0 ,x_1 ) = (t,x),\,\,0 < x < L, -
\infty  < x < \infty $), other conformal constant components of the
metric $ \eta ^1 ,\,\,\eta ^ \bot$ are unchanged external field.
From (\ref{4}), we get the Euler-Lagrange equation
\begin{equation} \label{18}
L_\varphi   = \eta ^ \bot  \Lambda e^\varphi   - \partial _0 \left[
{(\eta ^ \bot  )^{ - 1} \Omega } \right] - \partial _1 \left[ { -
(\eta ^ \bot  )^{ - 1} \Omega \eta ^1  - \eta ^ \bot  \varphi ' -
2\eta ^ {\bot  '}} \right] = 0,
\end{equation}
in which $ \Omega  = \dot \varphi  - \varphi '\eta ^1  - 2\eta^{1'}
$.

Now, let us consider Dirichlet boundary conditions
\[
\begin{array}{l}
 \delta \varphi (t,x = 0) = \delta \varphi (t,x = L) = 0, \\
 \delta \varphi '(t,x = 0) = \delta \varphi '(t,x = L) = 0 .\\
 \end{array}
\]

With these conditions, one part of the surface integrated term in
(\ref{3}) vanishes. Thus, for the equation of motion is satisfied,
since field function does not disappear at time boundary, it is
essential to supplement auxiliary conditions
\begin{equation} \label{19}
C(t,x) = \frac{{\partial L}}{{\partial \dot \varphi }} - \partial _i
\frac{{\partial L}}{{\partial \left( {\partial _i \,\dot \varphi }
\right)}} = 0,\,\,\,\,D(t,x) = \frac{{\partial L}}{{\partial \ddot
\varphi }} = 0.
\end{equation}
This is also an extension of Neuman boundary conditions.

It is obvious that energy-momentum tensor (\ref{6}) is modified by
relations (\ref{19}). Its new formulae in this case will be
\begin{equation} \label{20}
 T'_{\rho 0}  =  - \frac{{\partial L}}{{\partial \dot \varphi }}\varphi _{,\rho }  - \frac{{\partial L}}
 {{\partial \dot \varphi _{,\nu } }}\varphi _{,\rho \nu }  + \partial _0 \frac{{\partial L}}{{\partial \dot \varphi _{,\nu } }}
 \varphi _{,\rho }  + L\delta _{\mu \rho }, \,\,\, T'_{\rho 1}  = L\delta _{1\rho }.
\end{equation}

Using extended Neuman conditions (\ref{19}), we also have the new
surface integrated term (\ref{15})
\begin{equation} \label{21}
M = \frac{1}{2}\int\limits_R {\partial _1 \left( {\dot \varphi
\frac{{\partial L}}{{\partial \left( {\partial _i \dot \varphi }
\right)}}} \right)dx\,}  = \left. {\dot \varphi \frac{{\partial
L}}{{\partial \dot \varphi '}}} \right|_{x = 0}^{x = L}.
\end{equation}

Since Lagrangian (\ref{17}) does not depend explicitly on $\dot
\varphi '$, which means $\frac{{\partial L}}{{\partial \dot \varphi
'}} = 0$, the contribution of surface integrated term (\ref{21}) in
energy formula is vanished.

In summary, in $2$D gravity, using Dirichlet boundary conditions, it
is necessary to add extended Neuman conditions for the equation of
motion to be satisfied. Consequently, some minor changes appear in
energy-momentum tensor and surface integrated components does not
appear.

\section{STRING THEORY}
In the example of $2$D gravity, boundary condition of field function
was considered. The result is that when Dirichlet conditions are
applied, surface integrated terms vanish. In this section, for
string theory, boundary conditions of space-time coordinates are
studied.

The Lagrangian for the rigid string has the form
\begin{equation} \label{22}
L = \sqrt { - g} \;\left( { - \gamma  + \alpha \Delta x^\mu  \Delta
x_\mu  } \right),
\end{equation}
where $ \alpha  \ne 0,\,\, - \gamma  > 0$ are constants, $\alpha$ is
a dimensionless constant characterizing for the curvature of the
word-sheet of the string. For $\alpha  = 0$, we would obtain the
usual Nambu-Goto string. $ \Delta$ is Laplace-Beltrami operator.

Let consider this Lagrangian in two dimensional parametrization $ x
= (\tau,\sigma ),\,0\;\le \;\sigma \; \le \;\pi ,\,-\infty  \prec
\;\tau \; \prec \;\infty$. For this parametrization, in formulae in
Section II, instead of making derivatives with respect to variable
x, the variables now are parameters $(\tau ,\sigma )$. Moreover,
derivatives with respect to $ \varphi$ are now transformed into
derivatives with respect to $x$.

In the virtual of the above assumptions, form (\ref{4}) and
\cite{ref12}, we have the equation of motion
\begin{equation} \label{23}
\begin{array}{l}
 \left( {\gamma  - \alpha \Delta x^\mu  \Delta x_\mu  } \right)\Delta x_\mu   + 2\alpha \left( {\Delta \left( {\Delta x_\mu  } \right) - g^{i\rho } x_{,\;i}^\nu  \;x_{\mu \,,j} \Delta \left( {\Delta x_\nu  } \right)} \right) -  \\
 \,\,\,\,\,\,\,\,\,\,\,\,\,\,\,\,\,\,\,\,\,\,\,\,\,\,\,\,\,\,\,\,\,\,\,\,\,\,\,\,\,\,\,\,\,\,\,\,\,\,\,\,\,\,\,\,\,\,\,\,\,\,\,\,\,\,\, - \,4\alpha g^{i\rho } g^{k\,z} \left( {\Delta x_\nu  } \right)_{,\,j} x_{,\,k}^\nu  \;\nabla _i x_{\mu \,,\,z}  = 0 \\
 \end{array},
\end{equation}
where $\nabla _i x_{\mu \;,\,z}$ denotes the covariant derivative of
the covariant vector $ x_{\mu \;,\,z}$; $z=0,1,$ defined with the
use of Christofell symbols for the metric. Equation (\ref{23}) is
rather complicated since it contains fourth order derivatives and
nonlinear components. For $\alpha = 0$, this equation will yield to
Nambu-Goto equation $\Delta x_\mu   = 0$.

In the case of open rigid string, we choose boundary conditions as
in \cite{ref12}
\begin{equation} \label{24}
\delta x_\mu  \left( {\tau ,\,\sigma } \right) = 0(NG),\,\,\,\delta
x_{\mu ,\;0} \left( {\tau ,\,\sigma } \right) = 0\,\,\,for\,\,\,\tau
= \tau _1 ,\;\tau _2 ,\,\,\sigma  \in \left[ {0,\;\pi } \right].
\end{equation}

Then, for the equation of motion (\ref{23}) is satisfied, the
auxiliary conditions which the field function need to obey are (the
procedure is the same as in section III)
\begin{equation} \label{25}
\begin{array}{l}
 B_\mu  (\tau ,\,\sigma  = 0) = 0,\,\,\,B_\mu  (\tau ,\,\sigma  = \pi ) = 0, \\
 C_\mu  (\tau ,\,\sigma  = 0) = 0,\,\,\,C_\mu  (\tau ,\,\sigma  = \pi ) = 0, \\
 \end{array}
\end{equation}
where
\begin{equation} \label{26}
\begin{array}{l}
 B_\mu  (\tau ,\,\sigma ) = \sqrt { - g} \left( {\gamma  - \alpha \Delta x^\mu  \Delta x_\mu  } \right)g^{1\,i} x_{\mu ,\,i}  + 2\alpha \sqrt { - g} \,\;g^{j\,k} x^\lambda  _{,\,i} x_{\lambda \;,j\,k} \Delta x_\mu   +  \\
 \,\, \,\,\,\,\,\,\,\,\,\,\,\,\,\,\,\,\,\,\,\,\,\,\,\,\, + 4\alpha \sqrt { - g\;} \Delta x_\sigma  \;x^\sigma  _{,\,i\,j} \,g^{1\,j} g^{i\,k} x_{\mu \,,\,k}  + 2\alpha \partial _0 \left( {\sqrt { - g} \;g^{0\;1} \Delta x_\mu  } \right) +  \\
 \,\,\,\,\,\,\,\,\,\,\,\,\,\,\,\,\,\,\,\,\,\,\,\,\,\,\,+2\alpha \partial _j \left( {\sqrt { - g} \,g^{1\,j} \Delta x_\mu  } \right), \\
 \end{array}
\end{equation}
and
\begin{equation} \label{27}
C_\mu  (\tau ,\,\sigma ) = 2\alpha \sqrt { - g} \;g^{1\,1} \Delta
x_\mu.
\end{equation}

The energy density momentum $ p_\mu$ corresponding to Lagrangian
(\ref{22}) has the form
\begin{equation} \label{28}
\begin{array}{l}
 p_\mu = \sqrt { - g} \;g^{0\,j} \left( {\gamma  - \alpha \Delta x_\sigma  \Delta x^\sigma  } \right)x_{\mu \,,\,j}  + 2\alpha \partial _0 \left( {\sqrt { - g} \;g^{00} \Delta x_\mu  } \right) +  \\
  \,\,\,\,\,\,\,\,\,\,\,\,\, + 2\alpha \sqrt { - g} \;g^{0\,i} g^{j\,k} \left( {2\Delta x_\sigma  x^\sigma  _{,\,i\rho } \;x_{\mu \,,\,k}  + x^\lambda  _{,\,j\,k} \;x_{\lambda \,,\,i} \Delta x_\mu  } \right). \\
 \end{array}
\end{equation}

From (\ref{13}), (\ref{14}), (\ref{15}), we obtain the rate at which
the energy change with respect to the change in time \cite{ref12}
\begin{equation} \label{29}
\frac{{dH}}{{dt}} = the\;boundary\;terms =  - \partial _0 \left(
{\int\limits_0^\pi  {d\sigma } \partial _1 \left( {\mathop {\vec
x}\limits^ \bullet  \frac{{\partial L}}{{\partial \mathop {\vec
x}\limits^ \bullet  }}} \right)} \right).
\end{equation}

In general, the r.h.s of equation (\ref{29}) does not vanish. Thus,
conserved energy derived from energy-momentum tensor differs from
Hamilton function a surface integrated term. This also means the
evolution with time of Hamilton function of the rigid string is not
equal to its conserved energy

\section{CONCLUSION}
Classical equations of motion, energy-momentum tensor and angular
-momentum tensor of higher order derivative Lagranigan are
systemized and rederived in the appearance of boundary condition. We
have also pointed out that whether surface integrated terms can be
ignored or not depends on the boundary conditions. Two kinds of
boundary conditions are considered. In $2$D gravity, we use
Dirichlet conditions of field function for those terms to vanish. On
the contrary, in open rigid string, when boundary conditions of
space-time coordinates are chosen, it is unable to ignore surface
integrated terms. Moreover, some changes appear in the
time-components of energy-momentum tensor.
\section*{ACKNOWLEDGMENT }
The author would like to thank Prof. Nguyen Suan Han for his
suggestion of the problem and many useful comments.

\end{document}